\documentclass[epsf,twocolumn,showpacs,preprintnumbers]{revtex4}
\usepackage{graphics}
\usepackage{graphicx}
\usepackage{dcolumn} 
\usepackage{bm}
\usepackage{epsfig}
\begin{document}
\title{Na$_x$CoO$_2$ in the $x \rightarrow 0$ Regime:
  Coupling of Structure and Correlation effects } 
\author{K.-W. Lee and W. E. Pickett} 
\affiliation{Department of Physics, University of California, Davis, CA 95616}
\date{\today}
\pacs{71.20.-b,71.20.Be,71.27.+a}
\begin{abstract}
 The study of the strength of correlations in Na$_x$CoO$_2$ is
extended to the $x=0$ end of the phase diagram where Mott insulating behavior
has been widely anticipated. 
Inclusion of correlation as modeled by the LDA+U
approach leads to a Mott transition in the
$a_g$ subband if U$\geq$U$_c$=2.5 eV.  Thus U smaller 
than U$_c$ is required to model
the metallic, nonmagnetic CoO$_2$ compound reported by Tarascon and coworkers.
The orbital-selective Mott transition of the $a_g$ state, 
which is essentially degenerate with the
$e^{\prime}_{g}$ states, occurs because of the slightly wider bandwidth
of the $a_g$ bands.
The metal-insulator transition 
is found to be strongly coupled to the Co-O bond length, due to 
associated changes in the $t_{2g}$ bandwidth, but the largest effects 
occur only at a reduced oxygen height that lies below the equilibrium
position.
\end{abstract}
\maketitle

\section{Introduction}
The unusual, correlated electron behavior of the Na$_x$CoO$_2$ (NxCO)
system that has been confirmed in the $x \approx 0.6 - 0.8$ 
range\cite{corr_refs} attracted
the initial interest in this layered transition metal oxide system.
The discovery that hydration of $x\approx 0.3$ samples results in
superconductivity\cite{takada,others} 
with T$_c$ = 4.5 K has suggested analogies to the
high temperature superconducting cuprates -- a quasi-two-dimensional
transition metal oxide displaying
superconductivity only within a certain doping range --
and greatly heightened the excitement in this system.  While several
workers have already embarked on modeling the superconducting 
mechanism and symmetry,\cite{kuroki,johannes,khal,moch,yamanaka}
there remains a much more fundamental issue: just how strong are 
correlation effects where superconductivity is observed?
It is  now recognized\cite{ucdprl} that 
normal state properties do not show
signs of correlated behavior around $x \sim 0.3$. 
However, correlation effects are easily evident in the 
$x \approx 0.6 - 0.8$ range: magnetic fluctuations
are observed, the high temperature susceptibility reflects local moments,
the linear specific heat coefficient is strongly enhanced above the
band value, and magnetic ordering occurs in the high end of this range.
None of this behavior is observed for $x\approx$ 0.3: susceptibility and
NMR data show there are no magnetic ions, and the linear specific heat
coefficient is not enhanced over the band value.  

Signs of correlation effects are particularly evident at $x$ = 0.5,
where a metal-insulator transition (MIT) occurs\cite{cava_half} 
around 50 K, and strong signs of
charge disproportionation 2Co$^{3.5+}\rightarrow$ Co$^{3+}$ + Co$^{4+}$
are seen followed by ordering of the Co$^{4+}$ spins.  
Possibilities of such behavior for $x\approx 0.3$
have been discussed\cite{UCD1,baskaran}
because strong on-site repulsion, geometrical effects, and commensurability
all should conspire to encourage ordering at $x$=1/3.  Nevertheless,
there has been no experimental indication of any such 
tendencies,\cite{timusk} with
the most direct conclusion being that correlation effects seem really 
minor in this regime.  One suggestion\cite{ucdprl,UCD2} 
is that the multiband character
of the $x$=1/3 region considerably reduces the effect of the on-site
Coulomb repulsion compared to strongly correlated phenomena observed 
in the $x \approx$ 2/3 region.

Several new evaluations of the cobalt oxidation state in hydrated NxCO
lend a new urgency to the issue of degree of correlation in the system.
Takada and collaborators, who discovered the superconductivity
originally,\cite{takada} subsequently revealed that in their hydrated 
materials the formal valence of Co was +3.4 rather than the +3.65 that
would be inferred from the $x$=0.35 concentration of Na.\cite{takada2}  
They identified
the oxonium ion (also called hydronium) H$_3$O$^{+}$, one product of
the hydration product, as being an additional dopant of the CoO$_2$ layer.
Corroborating conclusions have been reached by Milne et al.\cite{milne}
and Karppinen et al.\cite{karppinen}: their superconducting materials
have a mean Co charge state close to +3.42 rather than the +3.65-3.7
that would be the case if only Na doping were responsible.  Chen 
{\it et al.}\cite{chen} report that they can obtain the charge-ordered
Co$^{+3.5}$ phase via doping with a combination of Na$^{+}$
and H$_3$O$^{+}$.  The implication of these reports is that superconductivity
occurs in the strongly correlated regime ``$x$'' $>$ 0.5 rather than in the
weakly correlated half of the phase diagram.

There have been few theoretical studies of the $x$=0 limit (CoO$_2$),
because this limit is extremely difficult to reach
experimentally and there is very limited data.  
Most of the models\cite{scy1,scy2,marianetti} of 
correlated-electron behavior in this system
implicitly (or explicitly) assume that the Co
on-site Coulomb repulsion U is strong enough that the $x$=0 phase will
be a Mott insulator.  The electronic structure studies so far have taken
this viewpoint as well,\cite{zhang,zou}  by
applying LDA+U correlated band theory
with a large enough Hubbard 
repulsion strength U to model
the presumed insulating phase.
The most direct way for this to happen is for
the $e^{\prime}_g$ states somehow to remain occupied and out of the picture,
and for Mott physics to occur within the $a_g$ band alone, resulting
in Mott insulating behavior.  (The layered structure and the distorted
CoO$_6$ edge-sharing structure result in the symmetry lifting 
$t_{2g}\rightarrow a_g + e^{\prime}_g$.) 
On the face of it, there are two impediments to
this scenario: (1) what breaks the balance between the $a_g$
and  $e^{\prime}_g$ orbitals, both of which contain a substantial
fraction of holes when $x$ is small? and (2) since the
existing experimental information indicates that CoO$_2$ is a
nonmagnetic metal rather than a correlated insulator, is CoO$_2$ close to
a Mott insulating phase or not?.  

In this paper we present an investigation of the $x$=0 compound CoO$_2$,
using the correlated band theory (LDA+U) approach but accounting also
for the experimental information that is available.  Conventional
band theory in the local density approximation (LDA) 
gives a half metallic (or nearly so) ferromagnetic
ground state for the CoO$_2$ layer throughout the NxCO system; this is not the
correct picture for $x <$ 0.75 where there is no magnetic order (except
exactly at $x=0.5$).  
This lack of magnetic order makes NxCO for $x <$ 2/3 a candidate
for a fluctuation induced paramagnet.\cite{singh2} 
The appropriate value of U for this
system is not well established, but seems to be significantly 
$x$-dependent,\cite{ucdprl,UCD2} so
we will vary its value within realistic limits to ascertain its
effect, and attempt to be consistent with the available data.

\section{Experimental Information}
Whereas it had been thought previously that CoO$_2$ was too unstable
to synthesize, Amatucci, Tarascon, and Klein\cite{amatucci} found that 
use of a dry plastic battery technology could be used to produce CoO$_2$
powders reproducibly by deintercalation of Li from Li$_x$CoO$_2$.  This
end compound was found to have a single CoO$_2$ layer hexagonal cell
with $a$=2.82~\AA, $c$=4.29~\AA.   The oxygen height in the octahedrally
coordinated CoO$_6$ arrangement was not established, but the compound is
metallic and nonmagnetic.\cite{personal}   The metallicity perhaps explains
why no Jahn-Teller distortion is seen for this nominally $d^5$ ion:
the metallic behavior indicates that the Co$^{4+}$ and O$^{2-}$ formal
valences cannot be taken as seriously as if it were an insulator.

Subsequent work by Tarascon {\it et al.}\cite{tarascon} on the
Li$_x$Ni$_{1-y}$Co$_y$O$_2$ system, and higher resolution data, 
reproduced the CoO$_2$ material.  The lattice constants were 
slightly smaller, $a=2.80$~\AA, 
$c=4.25$~\AA, and the oxygen position was weakly constrained 
by the data to the 
range 0.17$<z_0<$0.23.  This structure of CoO$_2$ results in 
neighboring layers of charged ``O$^{2-}$'' ions which have been thought to be the
cause of the instability of CoO$_2$ (and certainly would seem to 
contribute).  The possibility of O-O bonding across the van der Waals
gap in CoO$_2$ has been discussed,\cite{tarascon} and is consistent with the
observed metallic character.

\begin{figure}[tbp]
\rotatebox{-90}{\resizebox{5cm}{7cm}{\includegraphics{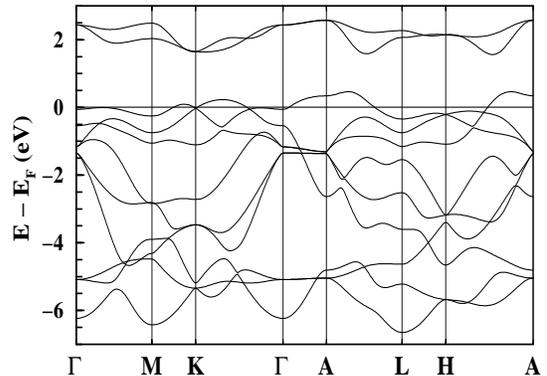}}}
\caption{LDA paramagnetic band structure of CoO$_2$.
  The e$_g$ bands lie above 1.5 eV, and the t$_{2g}$ bands in the
  range of -1 eV to 0.5 eV (but have strong hybridization with
  O 2p bands which are mainly below 1 eV). }
\label{pband}
\end{figure}

\section{Crystal Structure and Method of Calculation}
All reports\cite{tarascon,yang,venkatraman,jin} of CoO$_2$ 
indicate the familiar edge-sharing octahedron 
structure (space group P${\bar 3}$m1, No. 164). 
We used the experimental lattice constants a=2.8048 \AA, 
and c=4.2509 \AA,\cite{tarascon}
and optimized the oxygen height within LDA.
The resulting value is $z_0= 0.235c=0.999$ \AA, 
i.e. 95$^\circ$ Co-O-Co bond angle and 1.90 \AA~ Co-O bond length
(compare $z_0\simeq0.269c$ = 1.14~\AA, 90$^\circ$ and 1.98 \AA~
for undistorted octahedra).
In both FM and PM cases, the same relaxed value was found,
so the value is insensitive to magnetic order.
Venkatraman and 
Manthiram\cite{venkatraman} reported $z_0=0.257c$, but 
for nonstoichiometric CoO$_{1.72}$ samples.
Our calculated value is a 9\% larger than the relaxed value 
($z_0=0.908$ \AA, 98$^\circ$) at $x=$0.5 by Singh,\cite{singh1}
indicating a large effect of the Na$^+$ ions or the $c$ lattice
parameter, or perhaps as important, the added electronic 
charge.
When the lattice constants are fixed for all $x$, the relaxed O
height $z_0$ increases nearly monotonically 
for $x\geq 0.3$.\cite{zhang,johannes2} The optimized $z_0$ at $x=0$ 
by Zhang et al.\cite{zhang} is smaller by 0.1~\AA~  
than ours, because they used an artificially large $c$ lattice constant 
(50\% larger than the experimental one) to remove interlayer 
interactions.
The difference illustrates that the interlayer interaction 
({\it i.e.} effect of $c$ lattice constant) is significantly 
coupled to the O height.

\begin{figure}[tbp]
\rotatebox{-90}{\resizebox{6cm}{7cm}{\includegraphics{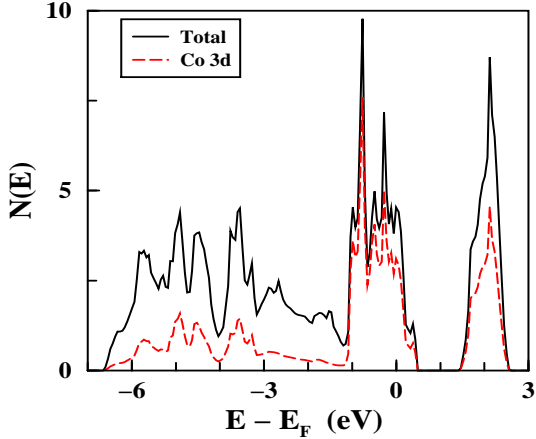}}}
\caption{(color online) LDA paramagnetic DOS of CoO$_2$.
   The crystal field splitting of t$_{2g}$-e$_g$ is 2.5 eV,
   and the bandwidth of the t$_{2g}$ manifold is 1.5 eV.}
\label{pdos}
\end{figure}

\begin{figure*}[tbp]
\resizebox{16cm}{6.5cm}{\includegraphics{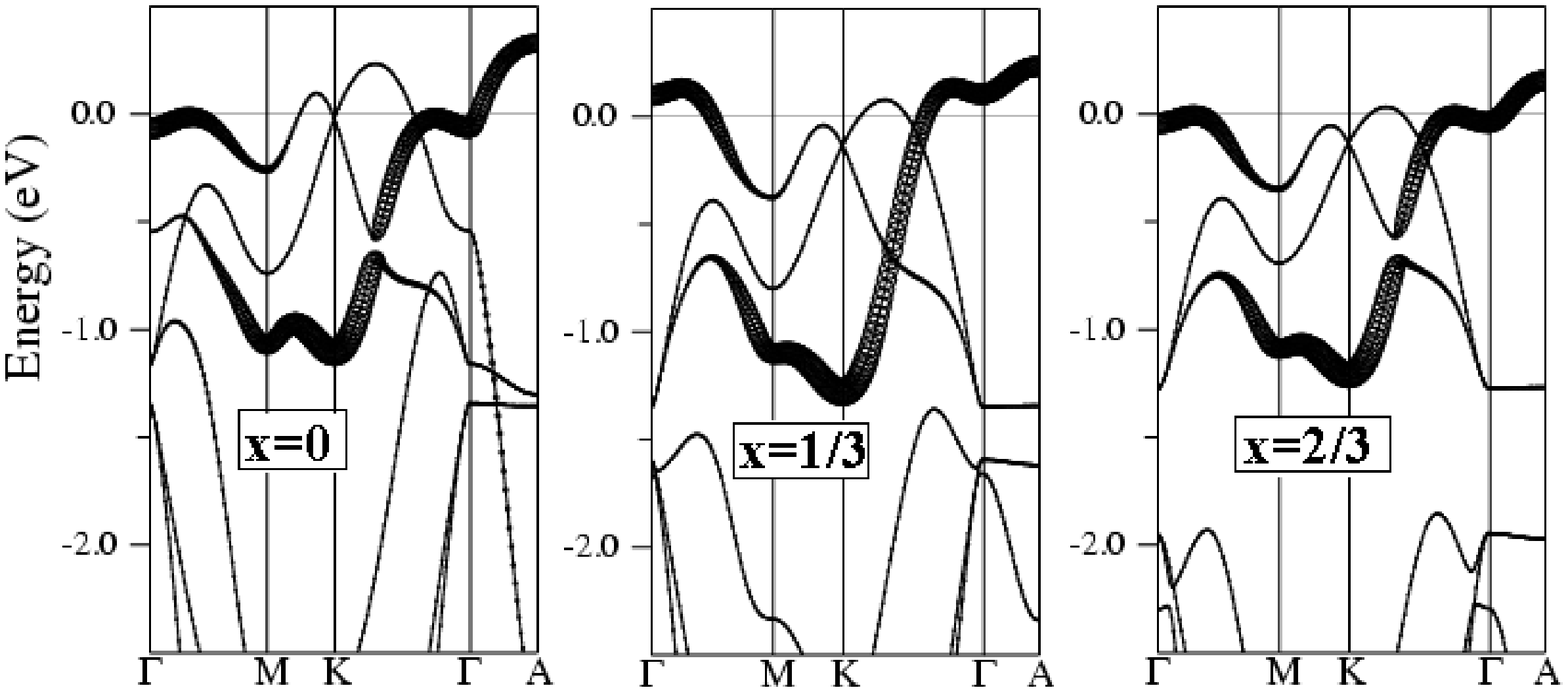}}
\caption{LDA paramagnetic blowup band structures near t$_{2g}$ manifold
  at $x=$0, 1/3, and 2/3 (in the virtual crystal approximation
  at $x=$1/3 and 2/3).
  As $x$ increases, the e$_{g}^\prime$ cylindrical Fermi surfaces are
  shrunk and the band separation (near -1.5 eV) along the $\Gamma$-A
  between t$_{2g}$ and O $2p$ becomes larger.
  The Fermi energy is at zero.
  The thickened lines emphasize the bands with strong a$_g$
  character.
  Lattice constants a=2.8048~\AA, c=4.2509~\AA~ and the oxygen
  height z$_0$=0.235 (0.999~\AA) are used at $x=$0, while a=2.84~\AA,
  c=5.405~\AA, and z$_0$=168 (0.908~\AA) at $x=$1/3 and 2/3.}
\label{xband}
\end{figure*}

The calculations reported here were carried out with two all-electron
full-potential electronic methods, full-potential local-orbital method
(FPLO)\cite{klaus} and Wien2k,\cite{wien}
with physically equivalent results.
Both popular schemes\cite{ldau1,ldau2}
for LDA+U functional were compared.
(The intra-atomic exchange integral $J$=1 eV was left unchanged.)
As found in previous studies\cite{ucdprl,UCD1,UCD2}, this cobaltate system
seems to lie in an intermediate regime where the two 
forms\cite{ldau1,ldau2} of LDA+U
functional give similar results; the difference in the 
double-counting corrections is small when the spin polarization is small.
The Brillouin zone was sampled with 162 (545 for the fixed spin moment
calculations) k-points in the irreducible wedge
and the basis sets were {\it 3s3p4s4p3d} for Co and {\it 2s2p3s3p3d} for O.
The full-potential linearized augmented-plane waves (FLAPW)
as implemented in Wien2k code\cite{wien} was also used
with the basis size determined by $R_{mt}K_{max}$=7.0 and APW sphere
radii (2.0 for Na and Co, 1.5 for O).
The calculations at $x=$1/3 and 2/3 were done with the virtual
crystal approximation (Na nuclear charge $Z=10 + x$) under
the same conditions as the previous calculations by the present
authors.\cite{UCD2}

\section{Nonmagnetic $\text{{\it x}=0: CoO}_{2}$}

This case corresponds most closely to nonmagnetic metallic CoO$_2$ as 
reported by Tarascon and coworkers.\cite{amatucci, tarascon}
The magnetic instability within LDA (incorrect, but generic for NxCO)
will be discussed below.
The LDA paramagnetic (PM) full band plot for CoO$_2$ is given 
in Fig. \ref{pband}, with
the corresponding density of states (DOS) given in Fig. \ref{pdos}.
The result is metallic, containing one hole per Co in the t$_{2g}$ bands.
The t$_{2g}$-e$_{g}$ crystal-field splitting is 2.5 eV, the same amount
as $x$=1/3, 1/2, and 2/3.\cite{UCD1,UCD2,singh1}
The t$_{2g}$ manifold is in the range of -1 eV to 0.5 eV relative to
the Fermi energy E$_F$, but is
hybridized with O $2p$ bands which are mainly below -1 eV, as can
be seen by the Co $3d$ character in the O $2p$ band region, and vice versa.

It is instructive to compare this $x=0$ limit with the $x > 0$ results
already published.\cite{ucdprl,UCD1,UCD2,zhang,zou,singh1,singh2,johannes2}
The variation with $x$ is provided for comparison in the enlarged
band structures around the t$_{2g}$ manifold at $x=$0, 1/3, and 2/3
given in Fig. \ref{xband}.
Although the Co-O bondlength at $x=$0 is 2.5\% larger than 
the $x=$1/3 and 2/3 values, suggesting less Co-O coupling,
in fact the $d-p$ mixing is strongest at $x=$0. 
The reason is that the decrease in mixing due to $t_{pd}$ 
is compensated by the decreasing in
the $\varepsilon_{t_{2g}}-\varepsilon_{p}$ energy separation.
The dispersion of the a$_g$ band along the $\Gamma$-A line
is much larger for $x$=0, so that quasi-two-dimensionality
is reduced as the Na content $x$ decreases (hole doping increases).
In addition, the band separation near -1.5 eV along the $\Gamma$-A line
between t$_{2g}$ and O $\it{2p}$ becomes larger.
The e$_{g}^\prime$ cylindrical Fermi surfaces containing
holes become smaller as $x$ increases,
and have disappeared at $x=$2/3,
consistent with previous reports.\cite{johannes2,zhang2}.
These changes contribute to the suggested
three-band to one-band crossover near $x\sim 0.5$.\cite{ucdprl,UCD2}

\begin{figure}[tbp]
\rotatebox{-90}{\resizebox{7cm}{8cm}{\includegraphics{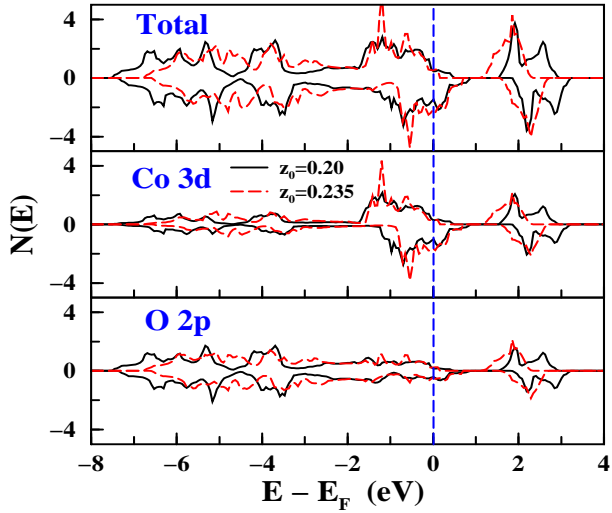}}}
\caption{(color online) Effect of the oxygen height $z_0$ on the LDA 
  ferromagnetic DOS at $x=$0. As the Co-O bondlength increases, 
  the t$_{2g}$-e$_g$ crystal-field splitting decreases, but the 
  mixing of Co $3d$ and O $2p$ remains strong.
   Note that it is not half-metallic, in contrast to the case of
   $0 < x < 1$, within LDA.
   The vertical dashed line indicates the Fermi energy.}
\label{zdos}
\end{figure}

\begin{figure}[tbp]
\rotatebox{-90}{\resizebox{8cm}{7cm}{\includegraphics{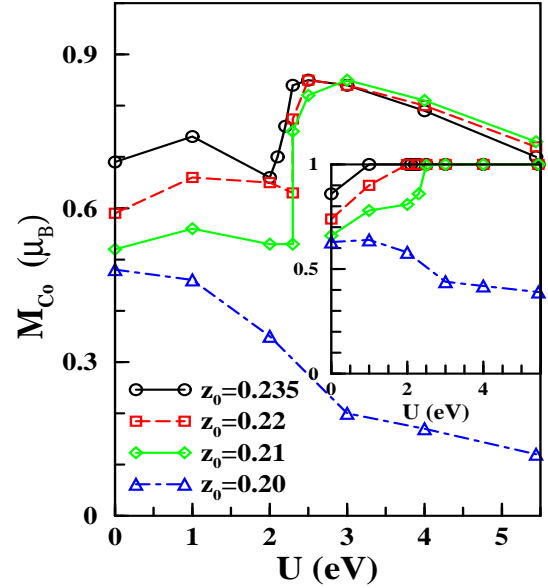}}}
\caption{(color online) Effect of the on-site Coulomb repulsion U 
  and the change of Co-O bondlength $z_0$ on magnetic moment 
  in FM state at $x=$0.
  At $z_0\leq 0.20$, the system is always metallic, regardless of U.
  (So it shows different behavior from $z_0\geq 0.21$.)
  At $z_0\geq 0.21$, there is a first order transition
  (metal to insulator) at U$_c$=2.5 eV. 
  Though $z_0$=0.235 case shows a little smooth transition,
  the sloop at U$_c$ is still very steep.
  The inset is change of the total magnetic moment with respect to U.}
\label{um}
\end{figure}

\section{Magnetic Tendencies}

\begin{figure}[tbp]
\rotatebox{-90}{\resizebox{8cm}{7cm}{\includegraphics{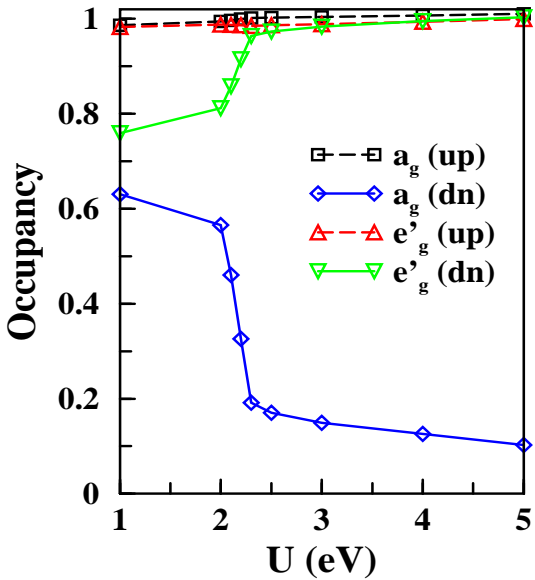}}}
\caption{(color online)Change of the occupancies of the a$_g$ and e$_g^\prime$ 
 states versus the on-site Coulomb repulsion U in FM state at the optimized 
 O height, which shows an important role of the a$_g$ band.
 The occupancy of the a$_g$ minority orbital decreases by 0.27$e$ at 
 U$_c$=2-2.5 eV, while both the majority orbitals are fully occupied 
 regardless of U.}
\label{occ}
\end{figure}

\begin{figure}[tbp]
\resizebox{7.5cm}{5.5cm}{\includegraphics{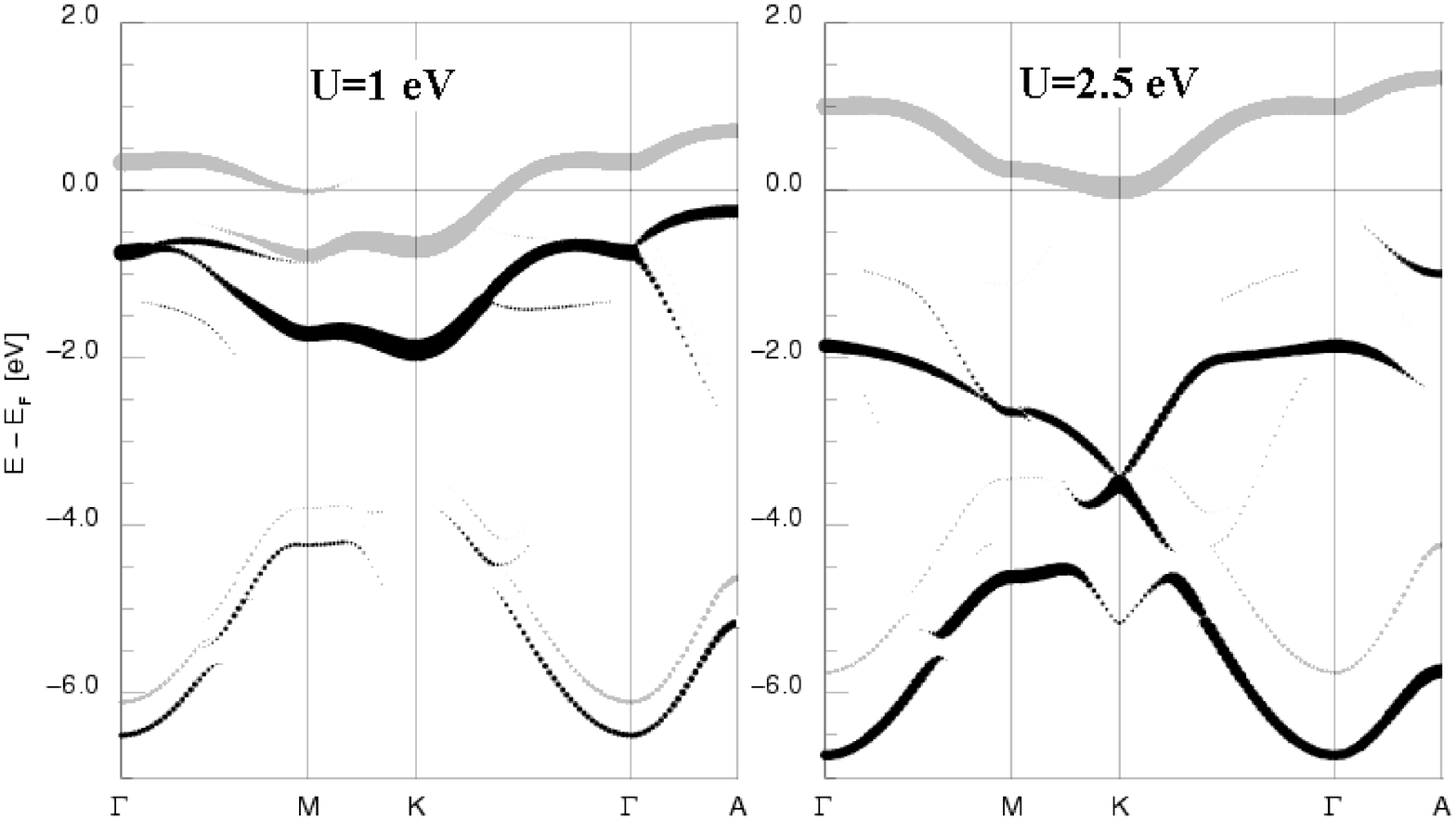}}
\caption{Change of the a$_g$ states versus the on-site Coulomb repulsion U 
 in FM state at the optimized O height,
 illustrating the Mott transition within the $a_g$ band.
 The gray lines indicate the minority character of the a$_g$ band, while
 the black lines denote the majority character.  The U=0 (LDA) result is
indistinguishable from the U=1 eV result shown here.}
\label{fat}
\end{figure}

As for all $0<x<1$,
a ferromagnetic (FM) state is favored energetically within LDA
at $x=$0 over a PM state.  The energy difference, 18 meV/Co, is quite 
small, considering the moment is near 1 $\mu_B$. 
For comparison, the energy gain due to a simple Stoner instability,
$Im^2/4$ with $I\approx$ 0.7-0.8 eV, would give a much larger value
approaching 150 meV.
The disagreement with experiments, which report 
no magnetism\cite{amatucci,tarascon,personal}, 
will be addressed in the next section.

Figure \ref{zdos} shows the ferromagnetic DOS at the relaxed height z$_0$=0.235, 
compared with z$_0$=0.20.
Increasing the oxygen height in this range (which relaxes the energy)
reveals three trends:
(1) narrowing of the unoccupied e$_g$ bands and the mostly occupied
 t$_{2g}$ bands,
(2) shift of the occupied O $2p$ bands toward E$_F$, and
(3) reduction of the crystal-field splitting.
In addition, as the Co-O bondlength, {\it i.e.} the O height $z_0$, increases, 
electrons transfer from O $2p$ to Co $3d$ at the rate
$\Delta Q_d/\Delta z_0 \approx$ 1.
The charge transfer is primarily into the a$_g$ majority band,
because the e$_{g}^\prime$ majority manifold is nearly fully occupied
already at  $z_0=$0.20.
At the relaxed $z_0$=0.235,  
the system is almost half-metallic.
As a related effect, the magnetic moment also increases 
as the O height increases. 

Change of the magnetic moment with the on-site Coulomb repulsion 
strength U is displayed in Fig. \ref{um} for four values of the 
oxygen height. 
At z$_0$=0.20, the magnetic moment
decreases rapidly above U=1 eV, since the system is still metallic
regardless of U. But at $z_0 \geq$ 0.21, the magnetic moments
jump sharply around U$_c$=2-2.5 eV, 
with the jump occurring simultaneously with
the gap opening. This first order transition was obtained also
at $x=$1/3, 1/2 and 2/3. \cite{ucdprl,UCD1,UCD2}
For $z_0 \geq$ 0.21, the amount of jump in the Co magnetic moment is 
decreased with increasing the O height. As a result, $z_0$=0.235 case 
shows a slightly smoothed transition, but the slope at U$_c$ is still very
steep.
This difference suggests that O height plays an important role in the 
metal-insulator transition.  For $U \approx U_c$, the oxygen modes may acquire
anomalous behavior due to coupling to the Co charge state.

The effect of the on-site Coulomb repulsion U on the occupancy of the
a$_g$ and e$_{g}^\prime$ states in Fig. \ref{occ} reflects the dominating
role of the a$_g$ minority band for the MIT, 
as it becomes the upper Hubbard band.
While the majority orbitals of the a$_g$ and e$_{g}^\prime$ states
are fully occupied regardless of U, the occupancies of the minority
orbitals show remarkable change at U$_c$=2-2.5 eV.
In particular, the occupancy of the $a_g$ minority orbital decreases 
by 0.27$e$ at U$_c$, leading to the Mott transition.
The transfer of $a_g$ spectral weight from U=1 eV to U=2.5 eV 
shown in Fig. \ref{fat}
vividly exposes the Mott character of the transition. 
At U$_c$=2.5 eV, the minority $a_g$ band is completely 
unoccupied while the majority band lies much deeper, mixing strongly
with the O 2$p$ bands and becoming spread over a 5 eV range.

In contrast to our previous findings for $x=$1/2,\cite{ucdprl} 
we have not been able to trace any hysteresis.  A hysteretic region
indicates that the two states are nearly degenerate, so the lack of
hysteresis indicates that the
difference in energy slopes $dE/dU$ of the two states differs much more than
for $x > 0$ case.

One other feature can be noted: the peak at U=1 eV in Fig. \ref{um} 
results from 
a transition from metal to half metal. 
The inset of Fig. \ref{um}, showing change of the total magnetic moment
with respect to U, shows the metal$\rightarrow$half metal transition clearly
since the total magnetic moment must be 1 $\mu_B$ 
for a half metal or an insulator. 
It is not completely clear why no Mott/disproportionation 
transition occurs at $z_0=0.20$.
However, it may be due to the increased 
hybridization between Co a$_g$ states and
O $p_z$ states. 
Even within LDA, the width of the a$_g$ bands at $z_0=0.20$ is
about 25\% larger than at $z_0=0.235$.
The increased hybridization will make the $a_g$-derived Wannier orbital 
less localized and therefore less susceptible to correlation effects.

\begin{figure}[tbp]
\rotatebox{-90}{\resizebox{7cm}{7cm}{\includegraphics{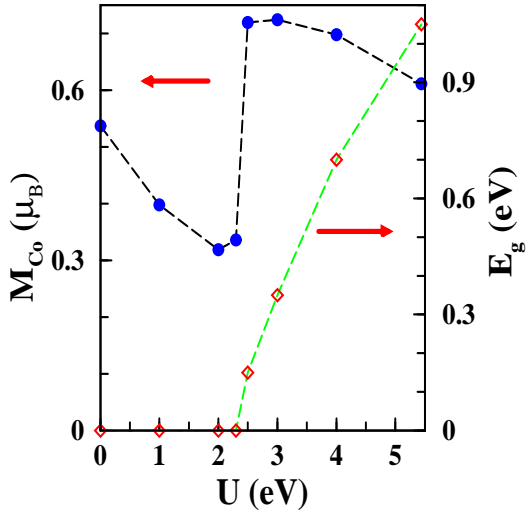}}}
\caption{(color online) Effect of the on-site Coulomb repulsion U on 
  magnetic moment in AFM state at $x=$0.
  There is a first order transition (metal to insulator) at U$_c$=2.3 eV.
  The right side indicates Co magnetic moment, and the left side does
  energy gap.}
\label{afm_um}
\end{figure}

{\it Antiferromagnetic tendency.}
 We used a two Co supercell with the optimized O height to study
 antiferromagnetism (AFM) (although FM is always favored 
energetically over AFM). 
The effect of the on-site Coulomb repulsion U on the magnetic moments
is shown in Fig. \ref{afm_um}.
In the region below U$_c$, the Co magnetic moment decreases rapidly as
U is increased, being only 60\% of LDA
at U=2 eV, a trend that is strongly related to Co$\rightarrow$O charge
transfer.
As for the FM case, there is a first order MIT at U$_c$= 2.3 eV, 
with the change of magnetic
moment (to 0.72 $\mu_B$) being even more dramatic than for FM case.

\section{Fixed Spin Moment Study}

\begin{figure}[tbp]
\rotatebox{-90}{\resizebox{6cm}{6cm}{\includegraphics{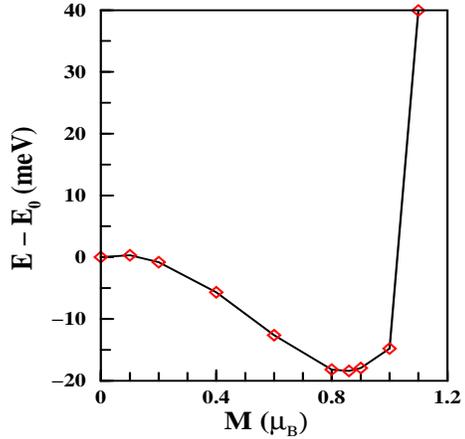}}}
\caption{LDA fixed spin moment calculation at $x=$0. 
  Note that a minimum occurs at $M$=0.86 $\mu_B$ before a sharp jump 
  at $M$=1 $\mu_B$. $M$ is the total magnetic moment per formula unit 
  in $\mu_B$. }
\label{fsm}
\end{figure}

As noted in the Introduction, CoO$_2$ is reported to be metallic 
and nonmagnetic. Both LDA and LDA+U calculations show energetically 
favored ferromagnetic solutions, as happens throughout the
NxCO system.\cite{UCD1,UCD2,singh1}
Fixed spin moment (FSM)\cite{schwarz} calculations help to  
establish the extent of
the contradiction. Figure \ref{fsm} shows the variation of total energy
with total magnetic moment on a per formula unit basis. 
The behavior is similar to the FSM results of Singh at $x=$0.3, 0.5, 
and 0.7.\cite{singh2} In particular, the sharp energy increase above 
$M=$1.0 $\mu_B$ is due to the crystal field induced gap between 
the $t_{2g}$ and $e_g$ manifolds.\cite{singh2}
However, the energy change vs. magnetic moment shows a slightly different
behavior from Singh's $x>0$ results. 
It has a minimum at $M$=0.86 $\mu_B$ (the value from the self-consistent 
calculation, close to but not quite half metallic) before increasing,
followed by the sharp jump
in energy at $M$= 1 $\mu_B$, which occurs upon filling the majority bands.

We have evaluated the Stoner (exchange interaction) I 
from these FSM calculations.\cite{rosner} 
This method properly accounts for the small contribution from the 
oxygen sites. The enhanced susceptibility is given by
\begin{eqnarray}
   \chi=\chi_{0}/[1-N(E_F)I]
\end{eqnarray}
where the bare susceptibility is given by $\chi_0=2\mu_{B}^{2}N(E_F)$ and 
N(E$_F$) is the single-spin
density of states at the Fermi level.
At small $M$, the energy difference is 
$E(M)-E(0)=(1/2)\chi^{-1}M^2$.
Since $N(E_F)$ for FM is 1.36 states/eV-spin, $I=$0.89 eV, [$I N(E_F)$ = 1.2]
comparable with the value obtained from the exchange splitting.


\section{Discussion and Summary}

In this paper we have explored the electronic structure and magnetic
tendencies of Na$_x$CoO$_2$, and specifically addressed the limits
on correlation that the experimentally reported nonmagnetic metallic
phase at $x$=0 imposes.
Taken in conjunction with results in the literature,
our study of the electronic structure and consequences of including
correlation effects in the band structure of CoO$_2$ extends the 
studies of NxCO across the entire $0 \leq x \leq 1$ regime.
Within this system,
increasing the Na concentration $x$ leads to three main effects.
First, the O $2p$ manifold 
moves downward from the $t_{2g}$ complex. 
This shift can be due to change in the Madelung potential, and 
possibly also to rehybridization of Co $3d$ and
O $2p$.\cite{marianetti} 
Secondly, the width of the $t_{2g}$ manifold shows some minor change with
increasing $x$, becoming a little narrower.  This can be interpreted as the
active $3d$ states becoming in effect more localized. 
Such a trend is consistent with our earlier suggestion  
that the on-site Coulomb repulsion U increases appreciably with
$x$.\cite{UCD2}
Thirdly, the band separation (between Co t$_{2g}$ and O $2p$) 
around -1.5 eV becomes obvious and increases as $x$ increases.
This trend reflects the fact that mixing between Co $3d$ and 
O $2p$ states becomes weaker as $x$ 
increases.  
Figure \ref{zdos} reveals that change of the Co-O bond 
length at $x=$0 hardly affects the separation, 
implying that the separation is induced by increasing 
electrons (or decreasing holes) as $x$ increases.

The observation by Tarascon and coworkers that CoO$_2$ is metallic
and nonmagnetic provides us with our
main conclusion: the effective Coulomb repulsion strength $U$ must be
less than $U_c = 2.3$ eV.  Above that value a Mott-type transition
occurs to a magnetic insulating phase.  This critical value is lower
than what has been obtained by the same methods for $x >$ 0, and 
supports the notion (and evidence) that the value of $U$ varies 
through the NxCO system, by at least a factor of two.

This value of $U_c$ in fact is
an upper limit; if the system were given more flexibility ({\it i.e.} if
we allowed antiferromagnetic ordering by choosing a supercell) some
ordering might occur for even smaller $U$.  The (unobserved) insulating
phase we find for $U > U_c$ corresponds to Mott transition of the $a_g$
orbital.  It is degenerate with the $e_g^{\prime}$ orbitals, however
its bandwidth is slightly wider.  As a result, allowing FM order even
within LDA already breaks the $a_g-e_g^{\prime}$ degeneracy by 
populating the $a_g$ orbital with most of the holes (see Fig. 7).  
Thus it is the
symmetry breaking arising from the slightly larger $a_g$ bandwidth that
results in the Mott transition of the $a_g$ states for $U > U_c$.
This choice is a type of orbital-selective Mott transition, which is
more commonly discussed at half-filling of the multiband 
system\cite{selective1,selective2} and thereby leaves metallic bands.  
Alternatives to a Mott transition of the $a_g$ state are possible, with  
a Mott transition within the $e_g^{\prime}$ orbitals coupled with
orbital ordering and accompanying symmetry-lowering distortion being one.
The (slightly) smaller bandwidth of the $e_g^{\prime}$ orbitals would be
a factor that would favor this possibility.

Finally, we remind that recent evidence\cite{takada2,milne,karppinen,chen}
suggests the oxidation state of Co for the superconducting materials is
representative of an effective doping level $x_{eff}$ = 0.55-0.60.  If true,
this puts superconductivity in a more interesting part of the phase
diagram, where superconductivity would arise from electron doping of
the $x$=0.5 disproportionated, charge-ordered, and spin-ordered magnetic
insulator.

\section{Acknowledgments}

We acknowledge helpful communications with V. I. Anisimov,
M. D. Johannes, J. Kune\v{s}, R. T. Scalettar,
D. J. Singh, R. R. P. Singh, and
J. M. Tarascon.
This work was supported by DOE grant DE-FG03-01ER45876 and DOE's
Computational Materials Science Network.
W. E. P. acknowledges support of the Department of Energy's
Stewardship Science Academic Alliances Program.

\end{document}